\documentclass[aps,pre,showpacs,twocolumn]{revtex4}
\usepackage{bm}
\usepackage{graphicx}
\bibstyle{approve.bib}
\usepackage{amssymb}
\usepackage{epstopdf}
\usepackage{verbatim} 
\usepackage{color}
\newcommand{\be}{\begin{equation}}
\newcommand{\ee}{\end{equation}}
\newcommand{\bea}{\begin{eqnarray}}
\newcommand{\eea}{\end{eqnarray}}
\newcommand{\lan}{\left\langle}
\newcommand{\ran}{\right\rangle}
\newcommand{\br}{\mathbf{r}}
\newcommand{\ba}{\mathbf{a}}

\newcommand{\bom}{\mathbf{\Omega}}

\newcommand{\e}{\varepsilon}

\newcommand{\tv}{\tilde{v}}

\newcommand{\pa}{\parallel}

\begin{document}

\title{Electrostatic interactions in charged nanoslits within an explicit solvent theory}

\author{Sahin Buyukdagli$^{1}$\footnote{email:~\texttt{Buyukdagli@fen.bilkent.edu.tr}}}
\affiliation{$^{1}$Department of Physics, Bilkent University, Ankara 06800, Turkey}
\date{\today}

\begin{abstract}
Within a dipolar Poisson-Boltzmann theory including electrostatic correlations, we consider the effect of explicit solvent structure on solvent and ion partition confined to charged nanopores. We develop a relaxation scheme for the solution of this highly non-linear integro-differential equation for the electrostatic potential. The scheme is an extension of the approach previously introduced for simple planes (S. Buyukdagli and Ralf Blossey, J. Chem. Phys. \textbf{140}, 234903 (2014)) to nanoslit geometry. We show that the reduced dielectric response of solvent molecules at the membrane walls gives rise to an electric field significantly stronger than the field of the classical Poisson-Boltzmann equation. This peculiarity associated with non-local electrostatic interactions results in turn in an interfacial counterion adsorption layer absent in continuum theories. The observation of this enhanced counterion affinity in the very close vicinity of the interface may have important impacts on nanofludic transport through charged nanopores.  Our results indicate the quantitative  inaccuracy of solvent implicit nanofiltration theories in predicting the ionic selectivity of membrane nanopores. 

\end{abstract}

\date{\today}
\maketitle

\section{Introduction}

Electrostatic interactions in confined liquids are relevant to various nanoscale processes ranging from macromolecular interactions~\cite{isr,netzvdw} to nanofiltration~\cite{yar1,yar2} and nanofluidic transport~\cite{Heyden2006,Stein}. An accurate theoretical formulation of these interactions has been a major challenge in the field of theoretical soft matter physics. The description of electrostatics in nanoscale systems has been mainly limited to Poisson-Boltzmann(PB) formalism~\cite{Gouy,Chapman}  known to suffer from two major limitations. First, the PB equation provides a mean-field (MF) description neglecting electrostatic correlations resulting from many-body interactions between charges.  The second weak point of the PB approach is the dielectric continuum approximation bypassing the extended charge structure of solvent molecules hydrating the ions of the solution.

In confined systems where the distance between charged macromolecules approach the Bjerrum length $\ell_w\approx7$ {\AA}, the above-mentioned limitations become drastic. First of all, the dielectric contrast between the macromolecule of low dielectric permittivity and the electrolyte of large permittivity is known to induce surface polarization effects. These non-MF effects that cannot be captured by the PB formalism are at the origin of i) attractive van-der-Waals forces stabilizing macromolecules~\cite{isr} and ii) dielectric exclusion phenomenon settling the salt selectivity of membrane nanopores in cells as well as in artificial nanofiltration devices~\cite{yar1}. Then, by ignoring the interaction between solvent molecules and the membrane, the PB formalism assumes the same solvent density and permittivity in the pore and the bulk reservoir. This dielectric continuum approximation is expected to result in two artefacts. First, solvent-membrane interactions are known to induce repulsive image-solvent effects leading to a reduced pore dielectric  permittivity~\cite{SBRB2}. This implies that solvent implicit electrostatic theories~\cite{pre1,jcp2} overestimate the pore permittivity and the strength of image-charge forces excluding ions from the nanopore. However, the same dielectric contrast between the pore and the reservoir should lead to an ionic Born energy barrier, i.e. an additional force favouring ionic rejection from the pore medium. Hence, the dielectric continuum limit giving rise to opposing artefacts is an uncontrolled approximation that has to be relaxed by the explicit consideration of the solvent charge structure.

Improvements over the mean-field and dielectric continuum approximations have been mainly considered separately. Within the dielectric continuum limit, electrostatic many-body effects have been included at one-loop~\cite{PodWKB,attard,netzcoun,1loop,SBRB3} and variational levels~\cite{netzvar,cyl1,Lue1,pre1,PRL,jcp2,SBTN}. In the MF approximation, explicit solvent theories have been developed by modeling water molecules as point dipoles~\cite{dunyuk,orland1}. The point dipole approximation being unable to account for non-local electrostatic interactions, we introduced in Ref.~\cite{NLPB1} the first solvent explicit formulation of non-local electrostatics with finite size dipoles. We explored next the MF non-linear response regime of this model in Ref.~\cite{SBRB1}. At this point, one should also mention the phenomenological non-local electrostatic theories developed in Refs.~\cite{Kor1,Kor2,Kor3,blossey1,blossey2}. Within the point dipole approximation, electrostatic correlations were considered in bulk liquids in Ref.~\cite{orland2} and at single charged planes in Ref.~\cite{epl} in order to scrutinize the effect of interfacial solvent configuration on the differential capacitance of low dielectric materials. 

The unification of non-locality and electrostatic correlations was introduced first in Ref.~\cite{SBRB2}. In this previous work, we derived the solvent-explicit variational equations with finite size dipoles. We solved these equations for an ion free solvent confined to neutral slits and also for an electrolyte in contact with a single charged interface. In the present article, we take a step further and develop a numerical scheme for the solution of the non-local PB equation (Eq.~(\ref{eq1}) in the main text) in charged membranes of slit geometry. In other words, the consideration of the liquid confinement and membrane charge is the novelty that had not been treated together in our previous article~\cite{SBRB2}. Within the framework of this formalism, we investigate the effect of surface charge and pore size on the configuration of solvent molecules, the liquid charge, and ion densities in the nanoslit. In particular, we show that due to an amplified surface field induced by the reduced solvent dielectric response at the interface, an ionic structure formation characterized by a double counterion concentration peak takes place. This structure formation corresponding to an enhanced counterion affinity is the most relevant result of the present work. We emphasize that the formalism developed herein provides the first explicit solvent theory of membrane selectivity.  The limitation of our theory and possible improvements are discussed in the Conclusion.

\section{Non-MF NLPB formalism}

\begin{figure}
\includegraphics[width=1.\linewidth]{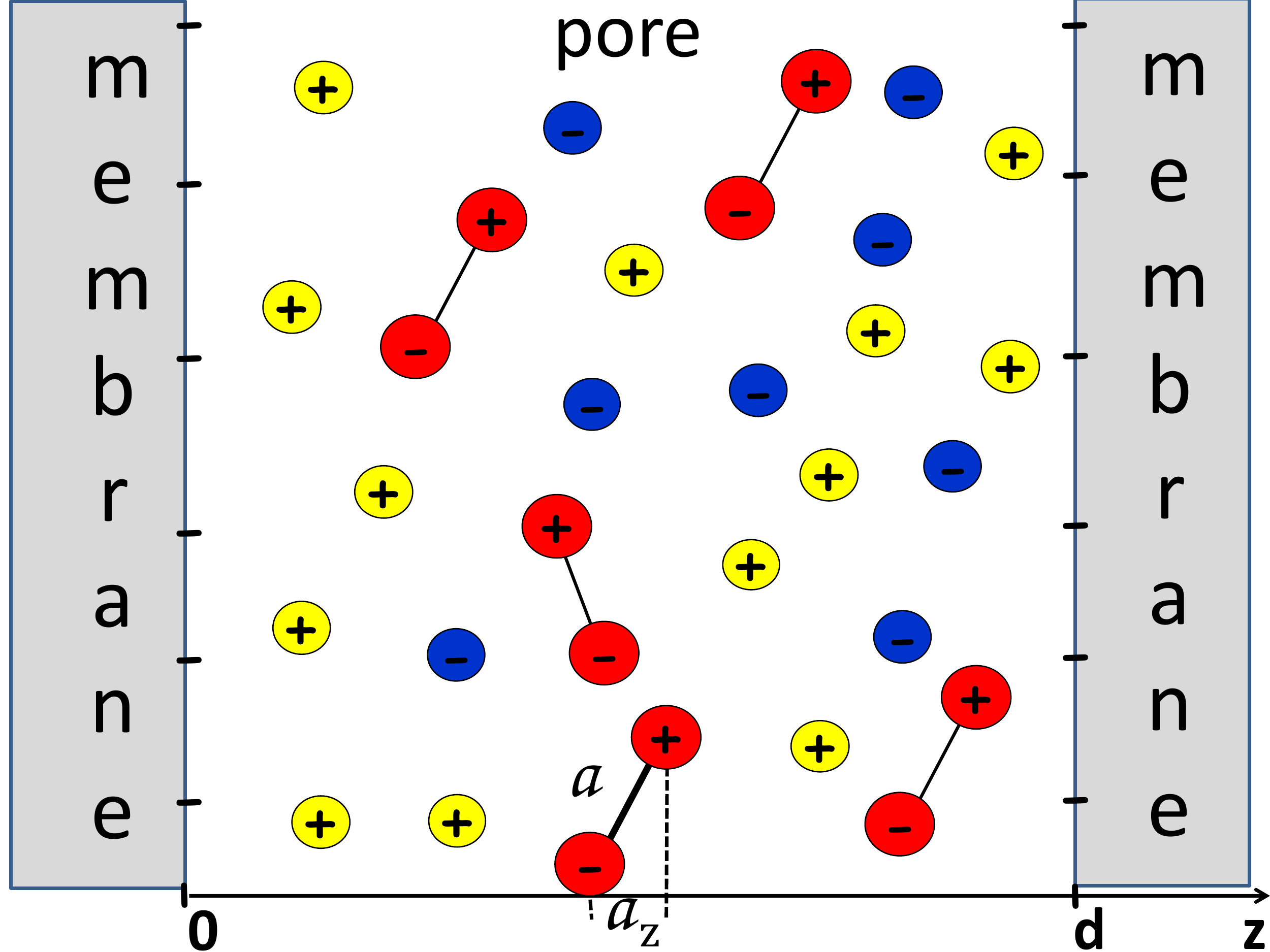}
\caption{(Color online) Schematic representation of the nanoslit with thickness $d$ and negative surface charge $-\sigma_s$. Dipoles of size $a$ (red) hydrate cations (yellow) and anions (blue). The dipolar orientation with respect to the $z-$axis is characterized by the projection $a_z$.}
\label{fig1}
\end{figure}

In this part, we review briefly the explicit solvent theory of electrolytes previously developed in Ref.~\cite{SBRB2}. The charged liquid is composed of solvent molecules and point ions confined to a slit pore. The composition of solvent molecules and the pore is depicted in Fig.~\ref{fig1}. The pore is composed of two infinitely large and negatively charged planes located at $z=0$ and $z=d$. The charged particles in the pore are in chemical equilibrium with an external charge reservoir. Each solvent molecule is modeled as a finite-size dipole composed of two oppositely charged point particles $\pm Q$ separated by the fixed distance $a$. The reservoir concentration of solvent molecules is $\rho_{sb}$. There are $p$ ionic species in the solution and each species $i$ has bulk concentration $\rho_{ib}$ with $i=1,...,p$. 

The beyond-MF NLPB equation for the electrostatic potential $\phi(z)$ in the solvent-explicit liquid was derived in Ref.~\cite{SBRB2}. This equation of state reads
\bea
\label{eq1}
&&\frac{k_BT}{e^2}\partial_z\e_0(z)\partial_z\phi(z)+\sum_iq_i\rho_i(z)+Q\left[\rho_{s+}(z)-\rho_{s-}(z)\right]\nonumber\\
&&=-\sigma(z),
\eea
where $k_B$ is the Boltzmann factor, $T=300$ K the ambient temperature, and $e$ the electron charge. The permittivity function is given by the piecewise form $\e_0(z)=\e_0\theta(z)\theta(d-z)+\e_m\left[\theta(-z)+\theta(z-d)\right]$, with $\e_0$ the vacuum permittivity and $\e_m$ the membrane permittivity. In the present work, dielectric permittivities will be expressed in units of the vacuum permittivity. Moreover, $q_i$ stands for the valency of the ionic species $i$, and the surface charge density reads $\sigma(z)=-\sigma_s\left[\delta(z)+\delta(d-z)\right]$. Finally, the ion number density and the density of the positive and negative charges on the solvent molecules are respectively given by 
\bea
\label{eq2}
\rho_{i}(z)&=&\rho_{ib}\;e^{-q_i\phi(z)-\frac{q_i^2}{2}\delta v_i(z)}\\
\label{eq3}
\rho_{s\pm}(z)&=&\rho_{sb}\int_{a_1(z)}^{a_2(z)}\frac{\mathrm{d}a_z}{2a}\;e^{-\frac{Q^2}{2}\delta v_d(z,a_z)}e^{\pm Q\left[\phi(z+a_z)-\phi(z)\right]}.\nonumber\\
\eea
In Eq.~(\ref{eq3}), the integration variable $a_z$ corresponds to the projection of the dipolar alignment vector $\ba$ onto the $z$ axis (see Fig.~\ref{fig1}) and the integration boundaries taking into account the rigidity of the interfaces are given by
\bea\label{eq4}
a_1(z)&=&-\mathrm{min}(a,z)\\
\label{eq5}
a_2(z)&=&\mathrm{min}(a,d-z).
\eea

We introduced in Eqs.~(\ref{eq2}) and~(\ref{eq3}) the renormalized ionic and dipolar self-energies $\delta v_i(z)$ and $\delta v_d(z,a_z)$ whose explicit form is reported in Appendix~\ref{ap1}. The relaxation algorithm that solves the integro-differential equation~(\ref{eq1}) is explained in detail in Appendix~\ref{ap2}~\cite{rem}.  We note that this algorithm generalizes to confined slits the scheme introduced in Ref.~\cite{SBRB2} for the single interface geometry. Furthermore, we note that in the present work, the charges composing the dipolar solvent molecules will be monovalent, that is $Q=1$. The solvent molecular size will be set to $a=1$ {\AA}, and the bulk solvent density will be taken as the liquid water density $\rho_{sb}=55$ M. Finally, we will consider exclusively the case of symmetric electrolytes composed of two monovalent ion species, i.e. $q_+=q_-=1$, each species with equal bulk densities $\rho_{+b}=\rho_{-b}=\rho_{ib}$.

\section{Results}
\subsection{Solvent configuration}

We investigate herein the partition of solvent molecules in the slit pore. To this aim, we will derive the number density and the average polarization of solvent molecules confined to the pore. We note that the dipolar part of the liquid grand-potential was derived in Ref.~\cite{SBRB2} in the form
\bea\label{eq6}
\Omega_d&=&-\rho_{sb}\int\mathrm{d}\br\frac{\mathrm{d}\bom}{4\pi}e^{-w_s(\br,\ba)}e^{-Q\left[\phi(\br)-\phi(\br+\ba)\right]}\\
&&\hspace{1.9cm}\times e^{-\frac{Q^2}{2}\delta v_d(\br,\ba)},\nonumber
\eea
where the multiple integral is taken over the spatial and orientational dipole configurations. Namely, the vector $\br$ indicates the position of the positive charge on the dipolar molecule. Then, the infinitesimal vector $\mathrm{d}\bom=\sin\theta\;\mathrm{d}\theta\;\mathrm{d}\varphi$ denotes the solid angle characterizing the dipolar orientation in the spherical coordinate system. The projection of the dipolar orientation vector onto the $z-$axis is related to the azimuthal angle as $a_z=a\cos\theta$ (see Fig.~\ref{fig1}). Moreover, the dipolar potential in Eq.~(\ref{eq6}) is $w_s(\br,\ba)=w_+(\br)+w_-(\br+\ba)+w_c(\br+\ba/2)$, where the functions $w_\pm(\br)$ are wall potentials acting on the positive and negative charges of solvent molecules, and $w_c(\br+\ba/2)$ is coupled to their centre of mass. Redefining now the position vector by shifting the latter to the centre of the solvent molecule as $\br'=\br+\ba/2$, evaluating the density from the functional derivative $\rho_d(\br')=\delta\Omega_d/\delta w_c(\br')$, taking into account the planar symmetry of the system, and performing a second transformation $a_z\to t=z-a_z/2$ on the variable of the integral over dipole rotations, one gets the dipole number density in the form
\bea\label{eq7}
\rho_s(z)&=&\rho_{sb}\int_{t_1(z)}^{t_2(z)}\frac{\mathrm{d}t}{a}\;e^{-\frac{Q^2}{2}\delta v_d(t,2z-2t)}\\
&&\hspace{2.1cm}\times e^{Q\left[\phi(2z-t)-\phi(t)\right]},\nonumber
\eea 
with the integration boundaries
\bea\label{eq8}
t_1(z)&=&\mathrm{max}(0,z-a/2,2z-d)\\
\label{eq9}
t_2(z)&=&\mathrm{min}(2z,z+a/2,d).
\eea
The discrete form of Eq.~(\ref{eq7}) is reported in Appendix~\ref{ap2}.  We also note that in the present work, the ionic and dipolar potentials $\delta v_i(z)$ and $\delta v_d(z)$ in the above equations will be calculated within the Debye-H\"{u}ckel (DH) approximation. This point is also explained in Appendix~\ref{ap2}. 

\begin{figure}
\includegraphics[width=1.2\linewidth]{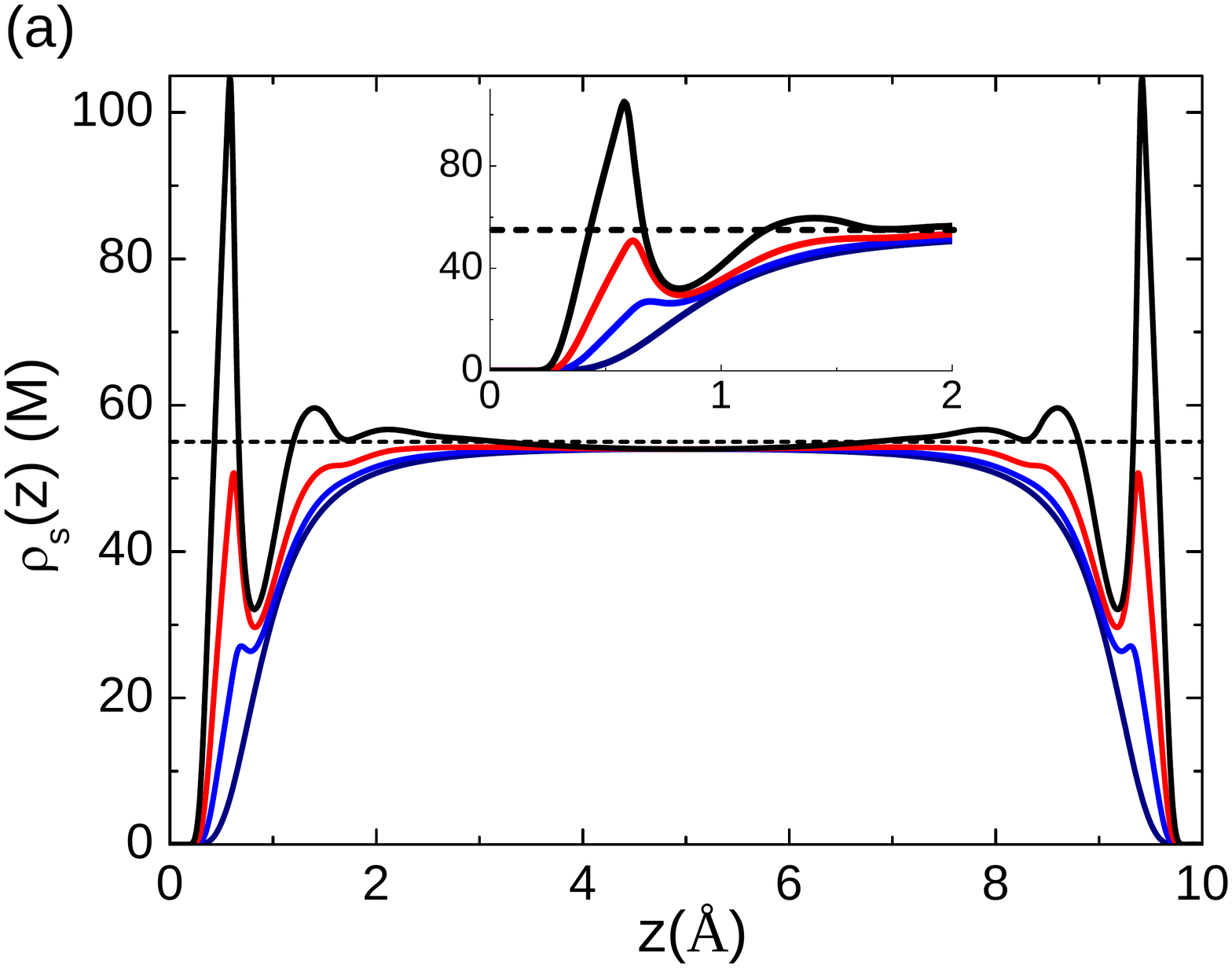}
\includegraphics[width=1.2\linewidth]{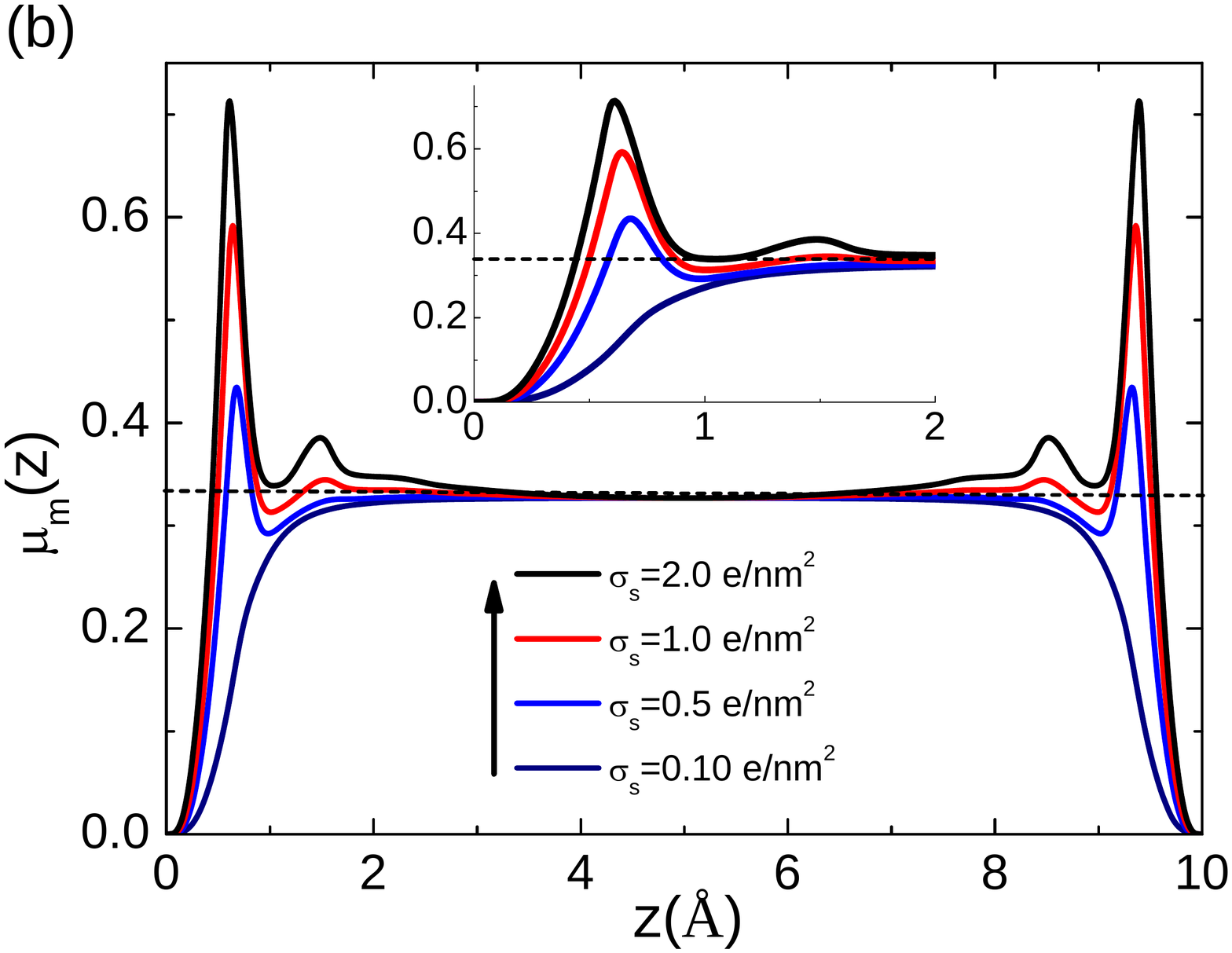}
\caption{(Color online) (a) Solvent number density~(\ref{eq7}) and (b) average polarization~(\ref{eq10}) for the charged solution confined in a slit of size $d=1$ nm and various surface charges displayed in the legend of the bottom plot. The membrane permittivity and salt concentration are respectively $\e_m=1$ and $\rho_{ib}=0.01$ M in both figures. The insets zoom on the interfacial region between $z=0$ {\AA} and $2$ {\AA}. The horizontal curves in (a) and (b) mark respectively the reservoir concentration and the limit of freely rotating dipoles $\mu_m(z)=1/3$.}
\label{fig2}
\end{figure}

We will show that solvent partition in the slit is driven by a competition between interfacial polarization and surface charge effects. The importance of each effect can be accurately described by the average dipole fluctuations~\cite{epl,SBRB2}. The latter can be expressed in terms of the number density~(\ref{eq7}) as
\be\label{eq10}
\mu_m(z)=\frac{\lan a_z^2\ran}{a^2\rho_d(z)},
\ee
with the statistical average
\bea\label{eq11}
\lan a_z^2\ran&=&\int_{t_1(z)}^{t_2(z)}\frac{\mathrm{d}t}{a}\;a_z^2\;e^{-\frac{Q^2}{2}\delta v_d(t,2z-2t)}\\
&&\hspace{1.3cm}\times e^{Q\left[\phi(2z-t)-\phi(t)\right]}\nonumber
\eea 
where $a_z=2(z-t)$. We note that the limit $\mu(z)=1/3$ corresponds to the case of freely rotating dipoles exclusively subject to entropy~\cite{epl,SBRB2}. This situation occurs in the bulk reservoir or far from the interfaces where image-dipole and surface charge-dipole interactions are absent. Then, in the regime $\mu_m(z)<1/3$ where repulsive image charge forces dominate surface charge effects, the former aligns dipoles parallel with the dielectric plane. This region is expected to be characterized by solvent depletion. Finally, for $\mu_m(z)>1/3$ where surface charge-dipole coupling overcomes image-dipole interactions, the charged interface attracts the positive charge and repels the negative charge of each dipole, resulting in the dipolar alignment perpendicular to the interface.  In this region, the surface charge should induce a local enhancement of the solvent density.

We illustrate in Figs.~\ref{fig2}(a) and (b) the number density and average polarization of dipoles in a slit of size $d=1$ nm. The model parameters are reported in the caption. The membrane permittivity set to $\e_m=1$ is characteristic of membrane nanopores. First, one notices that for intermediate to high surface charges, solvent density exhibits structure formation similar to the one observed in MD simulations~\cite{netznano1,netznano2}. Furthermore, the comparison of the top and bottom plots shows that the dipole density and the average polarization are strongly correlated, with the position of the minima and maxima coinciding almost exactly. First of all, in the top plot, one notes that for weakly charged membranes $\sigma_s=0.1$ e $\mbox{nm}^{-2}$, solvent molecules are depleted from membrane walls over the $3$ {\AA} thick layer.  Then, in the bottom plot and over the same region, the low values of the order parameter $\mu_m(z)<1/3$  indicates dipolar alignment parallel to the plane. Both the exclusion and parallel alignment effects are driven by the rotational penalty for dipoles as well as image-dipole interactions induced by the dielectric discontinuity at the interface. 

In fig.~\ref{fig2}(b), at the surface charge $\sigma_s=0.5$ e $\mbox{nm}^{-2}$, one sees that the average polarization exhibits a first peak above the free dipole limit. Thus, over this narrow region where surface-dipole interactions dominate image-dipole forces, solvent molecules show a tendency to align perpendicular to the charged wall. In the top plot and for the same surface charge, the amplified dipole-surface coupling is seen to yield as well a small peak in the density curve. Then, a further increase of the surface charge to the value  $\sigma_s=1.0$ e $\mbox{nm}^{-2}$ amplifies the first peak of the average polarization and dipole density curves, and it also gives rise to a second peak in the polarization. This corresponds to the surface charge regime where structure formation takes place. However, due to the presence of strong image forces in the confined pore, the solvent partition is still characterized by an overall exclusion from the pore, that is $\rho_d(z)<\rho_{sb}$. By setting the surface charge to the higher value  $\sigma_s=2.0$ e $\mbox{nm}^{-2}$, the pronounced surface charge attraction leads to three peaks in the density curve where solvent density exceeds the bulk value. The first two peaks are separated by a second solvent exclusion layer. Most importantly, at the position of the first peak, the solvent density reaches twice the reservoir concentration. We emphasize that such a strong interfacial enhancement of solvent number densities has been observed in MD simulations for both hydrophobic and hydrophilic surfaces (see e.g. Fig.1 of Ref.~\cite{netznano1}).

\begin{figure}
\includegraphics[width=1.25\linewidth]{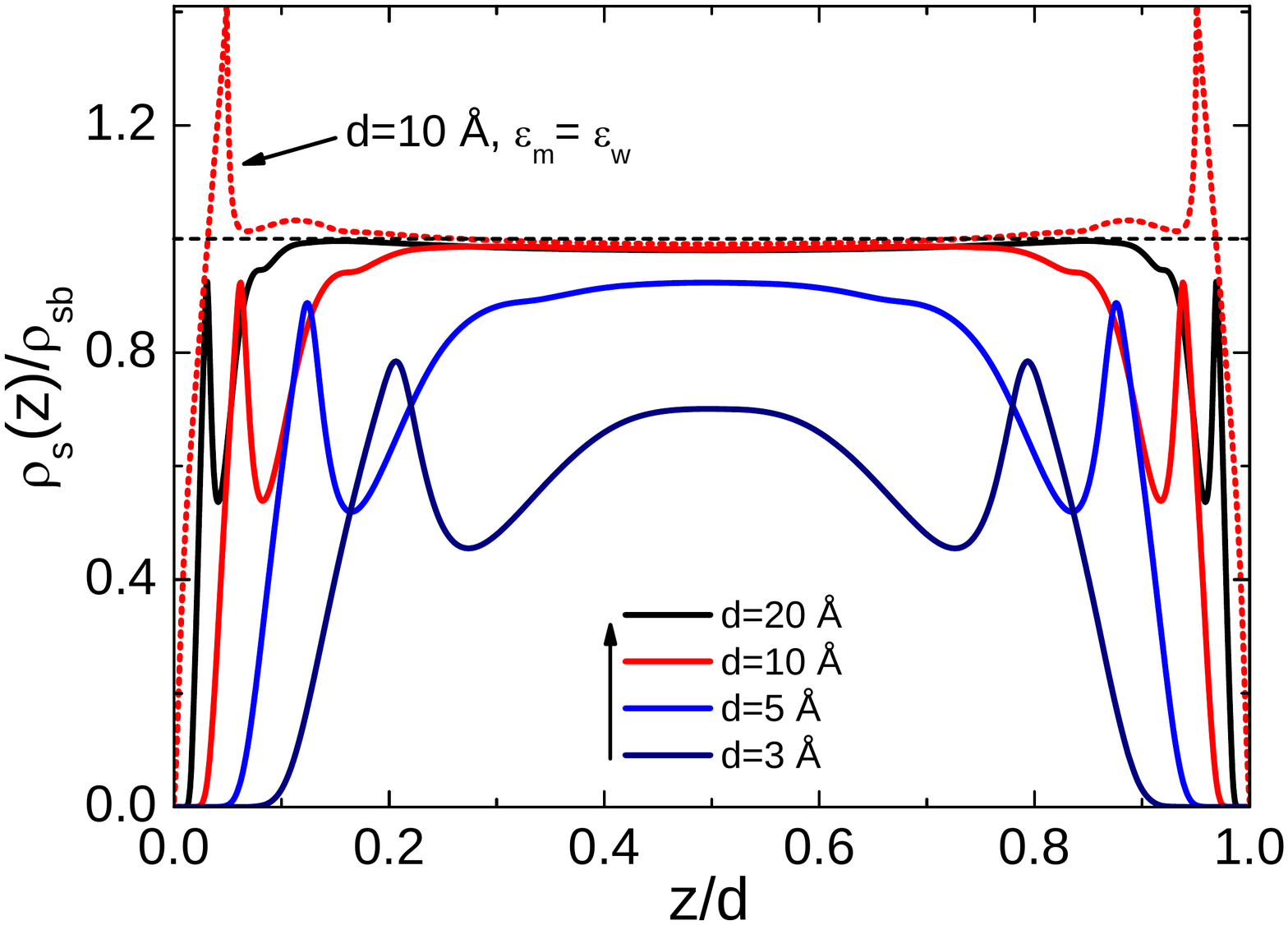}
\caption{(Color online) (a) Solvent dipole density rescaled with the bulk density for various pore sizes reported in the legend. The horizontal axis displays the position of the molecule rescaled by the pore size. The model parameters are the same as in Fig.~\ref{fig2}, with the surface charge $\sigma_s=1.0$ e $\mbox{nm}^{-2}$. Solid curves : $\e_m=1$. Dotted red curve : $\e_m=\e_w$.}
\label{fig3}
\end{figure}

In order to investigate the effect of confinement on solvent partition, we plotted in Fig.~\ref{fig3} the solvent number density for various pore sizes ranging between $d=3$ {\AA} and $20$ {\AA}. One notes that the decrease of the pore size that amplifies image-dipole interactions intensifies solvent exclusion from the pore. This effect is particularly noticeable below the nanometer pore size (i.e. for $d\lesssim\ell_w$) where the midpore density is significantly lowered. We emphasize that the strength of the dielectric exclusion effect determining the ionic selectivity of the membrane depends on the pore solvent density since the latter sets the intensity of the dielectric discontinuity between the membrane and the nanopore. Thus, the significant solvent rejection in narrow pores indicates that solvent-implicit nanofiltration models (see e.g. Refs.~\cite{yar1,yar2}) assuming the same solvent density in the pore and the reservoir are quantitatively inaccurate in predicting the ion rejection efficiency of subnanometer pores. In Fig.~\ref{fig3}, one also notes that the position of the first peak resulting from the competition between image-dipole forces, steric effects, and surface charge attraction stays weakly sensitive to the pore size. Indeed, multiplying the adimensional position of each peak with the corresponding pore size, one finds that the former is always located at $0.6$ {\AA}-$0.7$ {\AA}. 

Finally, in order to evaluate the importance of the dielectric discontinuity, we calculated the solvent density for a dielectrically continuous membrane (i.e. $\e_m=\e_w$ at the pore size $d=10$ {\AA}). One observes that in the absence of image-dipole interactions, the solvent decrement layer is replaced by an increment layer with $\rho(z)\geq\rho_{sb}$ for $z\geq0.5$ {\AA} $=a/2$. At $z=0.5$ {\AA}, we notice a sharp peak where steric effects resulting from finite solvent size $a=1$ {\AA} take over surface charge attraction and decrease the density of solvent molecules subject to rotational penalty below this point. The closeness of the peak positions for $\e_m=1$ and $\e_m=\e_w$ suggests that in explicit solvent MD simulations, the position of the first density peak is mainly determined by the finite solvent size rather than surface polarization effects. This point is supported by previous MD simulations  (see e.g. Figs.1(c) and (d) of Ref.~\cite{netznano1}) where the peak positions located at $2.5$ {\AA} - $4.0$ {\AA} for both hydrophobic and hydrophilic interfaces roughly correspond to the molecular size of the water TIP models. In the light of these results on solvent partition, we will scrutinize next the partition of the liquid charge in the nanoslit.

\subsection{Partition of the liquid charge}
\label{partch}

\begin{figure}
\includegraphics[width=1.2\linewidth]{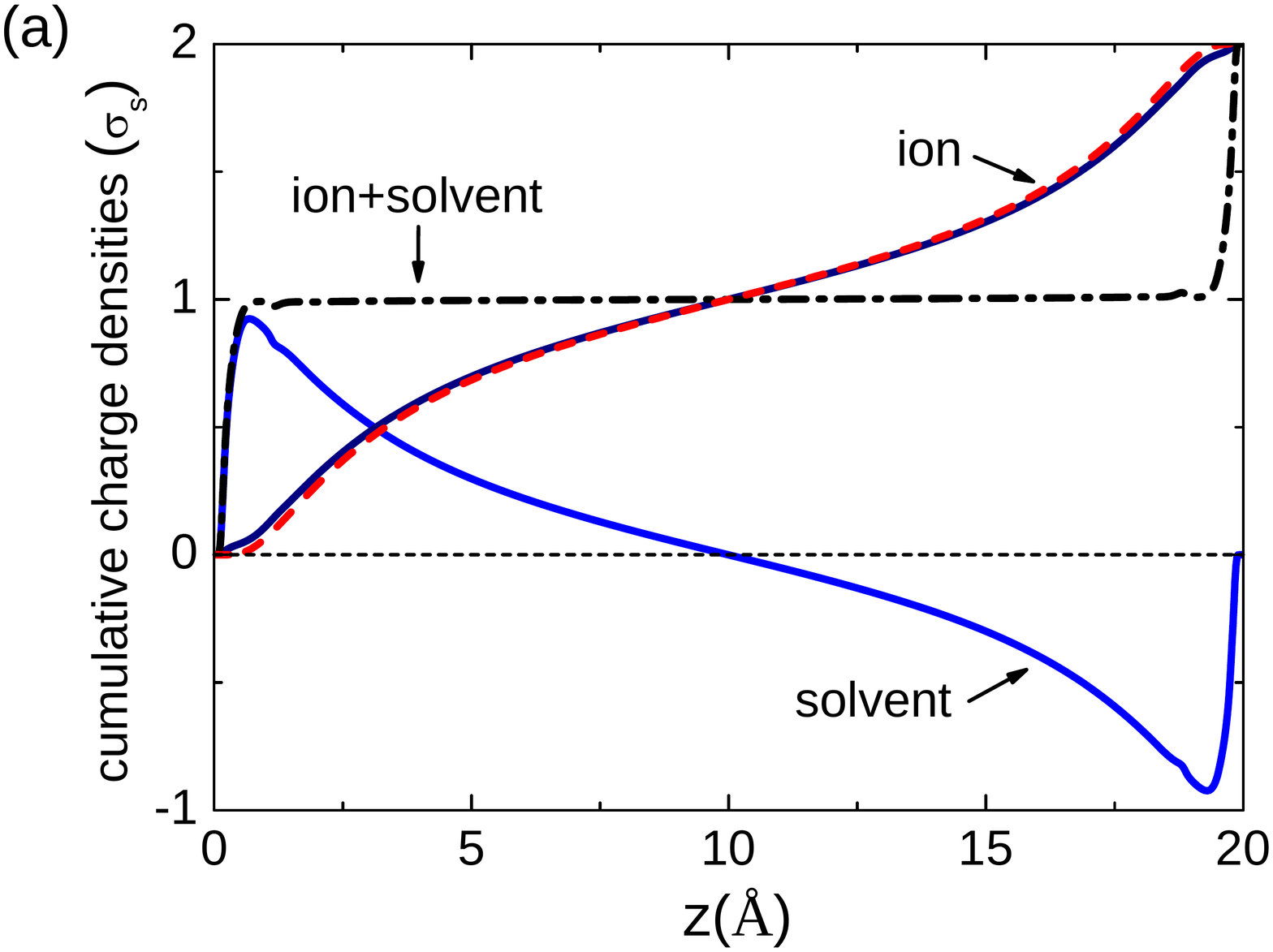}
\includegraphics[width=1.2\linewidth]{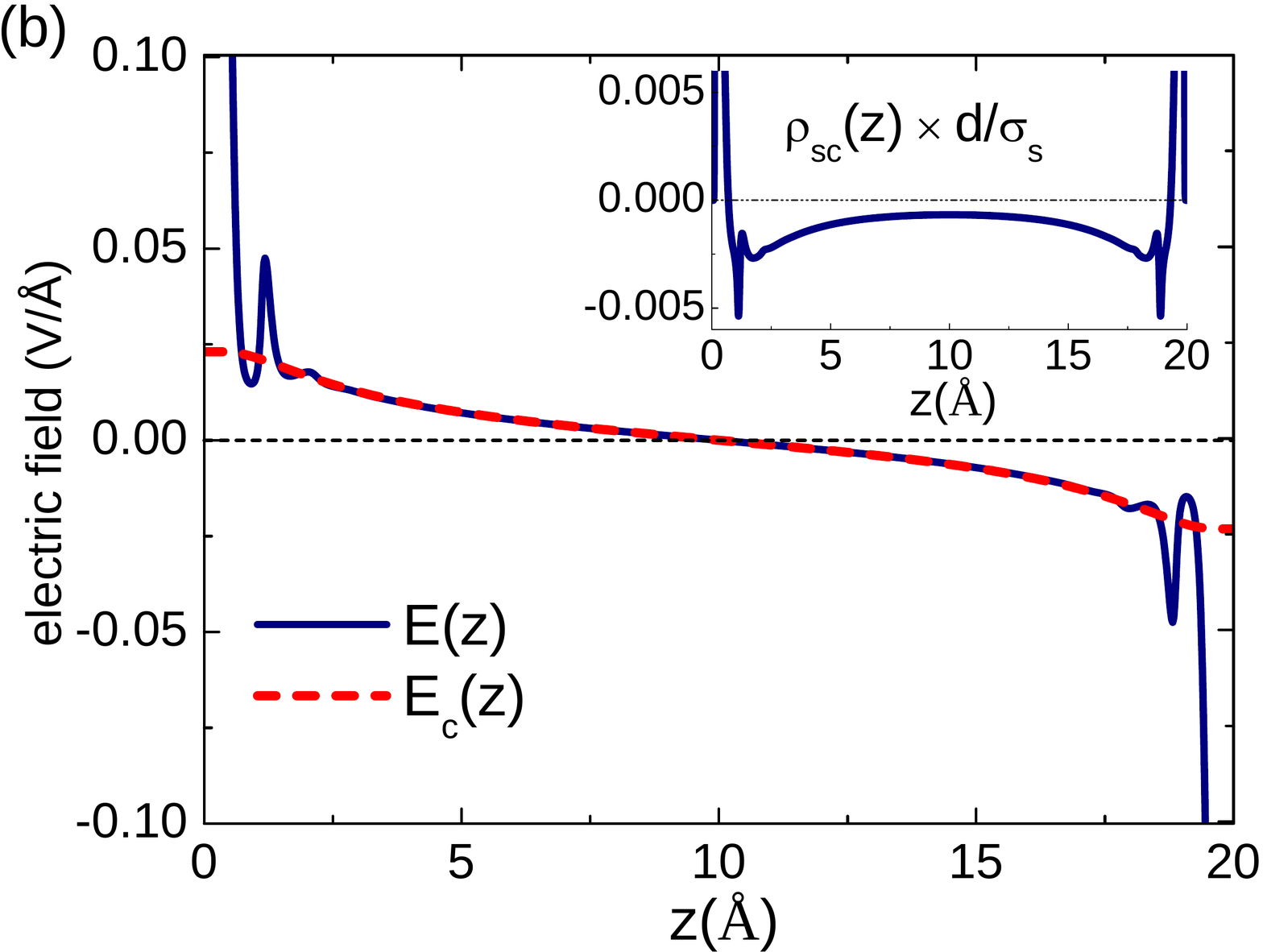}
\caption{(Color online) (a) Cumulative charge density of solvent molecules $\rho_{cum,s}(z)$ (solid blue curve), ions $\rho_{cum,i}(z)$ (solid navy curve), and total cumulative charge density $\rho_{cum,s}(z)+\rho_{cum,i}(z)$ (dashed-dotted black curve). The dashed red curve displays the cumulative ion density of the continuum theory. (b) Electric field from the explicit solvent (solid navy curve) and continuum theory (dashed red curve). The inset displays the local solvent charge density $\rho_{s+}(z)-\rho_{s-}(z)$. In (a) and (b), membrane permittivity is $\e_m=1$, bulk ion density $\rho_{ib}=10^{-2}$ M, membrane surface charge $\sigma_s=1.0$ e $\mbox{nm}^{-2}$, and pore size $d=2$ nm.}
\label{fig4}
\end{figure}

In order to investigate the charge partition in the nanoslit, we plotted in Fig.~\ref{fig4}(a) the cumulative ion density
\be
\label{eq12}
\rho_{cum,i}(z)=\int_0^z\mathrm{d}z'\left[\rho_+(z')-\rho_-(z')\right]
\ee
obtained from the solution of the explicit solvent PB equation~(\ref{eq1}) (solid navy curve) and the continuum result (dashed red curve) from the numerical solution of the modified PB equation~\cite{pre1}
\bea
\label{eq13}
&&\frac{kT}{e^2}\partial_z\e(z)\partial_z\phi_c(z)+\sum_iq_i\rho_i(z)=-\sigma(z),
\eea
with the sharp dielectric jump function $\e(z)=\e_w\theta(z)\theta(d-z)+\e_m\left[\theta(-z)+\theta(z-d)\right]$. The modified PB formalism ignores the non-uniform solvent partition in the pore. Fig.~(\ref{fig4})(a) shows that moving from the interface at $z=0$ towards the second interface at $z=d$, the cumulative ion density increases monotonically and reaches the surface charge value in the mid-pore $\rho_{cum,i}(d/2)=\sigma_s$ and twice the surface charge at the second interface, i.e. $\rho_{cum,i}(d)=2\sigma_s$. This is a consequence of the electroneutrality condition according to which there must be as many monopole charges in the nanoslit as the fixed surface charge on the membrane walls. One also notes that the cumulative ion density from the NLPB formalism (navy curve) and continuum theory (red curve) are very close to each other. Thus, solvent structure plays a minor role in the cumulative ion charge configuration. 

We now note that integrating Eq.~(\ref{eq13}) from the interface at $z=0$ to any point $z$ located in the pore, one can express the continuum electric field $E_c(z)=\partial_z\phi_c(z)$ as
\be
\label{eq14}
E_c(z)=4\pi\ell_w\left[\sigma_s-\rho_{cum,i}(z)\right].
\ee
We reported the electric field profile $E_c(z)$ in Fig.~\ref{fig4}(b). As suggested by Eq.~(\ref{eq14}) and in agreement with the ion density curve in Fig.~\ref{fig4}(a), the electric field evolves from $E_c(0)=4\pi\ell_w\sigma_s$ at the interface to zero in the midpore where $\rho_{cum,i}(d/2)=\sigma_s$ and the field reaches the value $E_c(d)=-4\pi\ell_w\sigma_s$ at the second interface where the cumulative ion density is twice as large as the wall charge. In other words, within the continuum formalism, the spatial variation of the field is solely determined by the ion screening effect. 

We consider next the cumulative solvent density
\be
\label{eq15}
\rho_{cum,s}(z)=\int_0^z\mathrm{d}z'\left[\rho_{s+}(z')-\rho_{s-}(z')\right]
\ee
displayed in Fig.~\ref{fig4}(a) by the solid blue curve together with the local solvent density shown in the inset of Fig.~\ref{fig4}(b). In the latter figure, one notes that the interfacial regions are each characterized by a positively charged solvent layer separated by a negative solvent layer. In Fig.~\ref{fig4}(a),  this local charge distribution is shown to result in an initial increase of the cumulative solvent density up to the maximum value $\rho_{cum,s}(z)\simeq\sigma_s$ where it starts to decrease and vanishes in the mid-pore. For $z>d/2$, the cumulative density becomes negative and drops monotonically up to the vicinity of the second interface located at $z=d$ where it reaches its minimum value. Moving closer to the second interface, due to the interfacial positive solvent charge layer, the cumulative density rises and becomes zero on the pore wall. The latter point shows that solvent molecules bring no contribution to the total charge in the slit. 

Before considering the total cumulative charge density, one notes that by integrating Eq.~(\ref{eq1}), one gets the equivalent of Eq.~(\ref{eq14}) for the electric field in the solvent-explicit liquid,
\be
\label{eq16}
E(z)=4\pi\ell_B\left[\sigma_s-\rho_{cum,i}(z)-\rho_{cum,s}(z)\right].
\ee
Eq.~(\ref{eq16}) indicates that the electric field profile is determined by the combined effects of salt screening (the second term on the right-hand-side) and dielectric screening associated with polarization charges (the third term). We plotted in Fig.~\ref{fig4}(a) the total charge density $\rho_{cum_i}(z)+\rho_{cum,s}(z)$ (dotted-dashed curve) and reported the electric field~(\ref{eq16}) in Fig.~\ref{fig4}(b) (solid navy curve). Observing both figures, one first notes that the interfacial solvent depletion responsible for a dielectric screening deficiency results in a surface field significantly larger than the field of the continuum theory. Then, due to the rise of the cumulative solvent density, the total cumulative charge density increases at the interface.  As a result of the amplified dielectric screening effect, the electric field decreases rapidly from the surface value $E(0)=4\pi\ell_B\sigma_s$ to the magnitude of the continuum field $E_c(z)$.  Moreover, we see that far away from the interfaces, the total cumulative charge density stays very close to the wall charge.  Finally, we note that the field $E(z)$ exhibits fluctuations around the continuum field. This arises from the solvent charge density fluctuations displayed in the inset of Fig.~\ref{fig4}(b). In the next part, we will investigate the effect of non-uniform solvent charge distribution on the partition of ions in the nanoslit.

\begin{figure}
\includegraphics[width=1.2\linewidth]{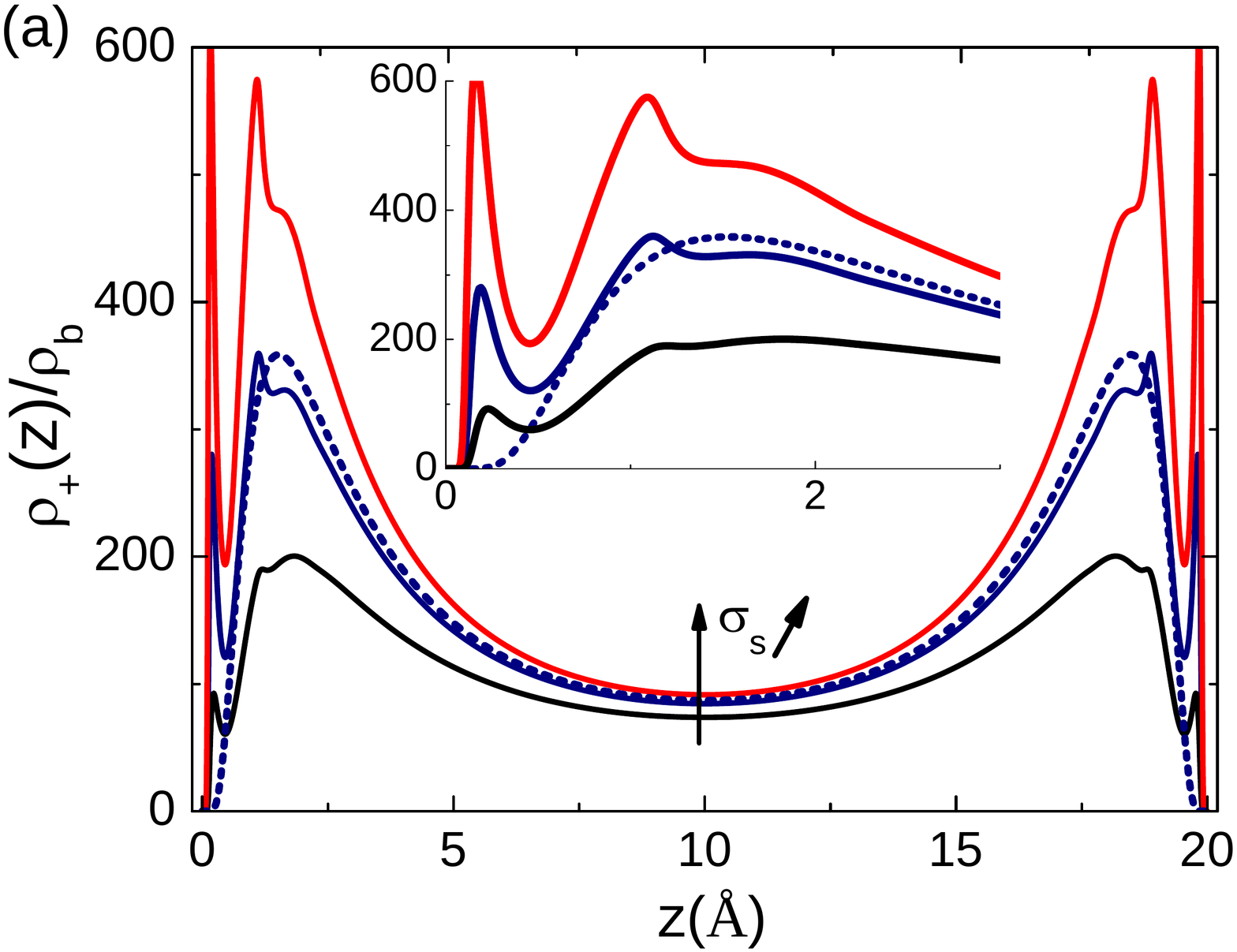}
\includegraphics[width=1.2\linewidth]{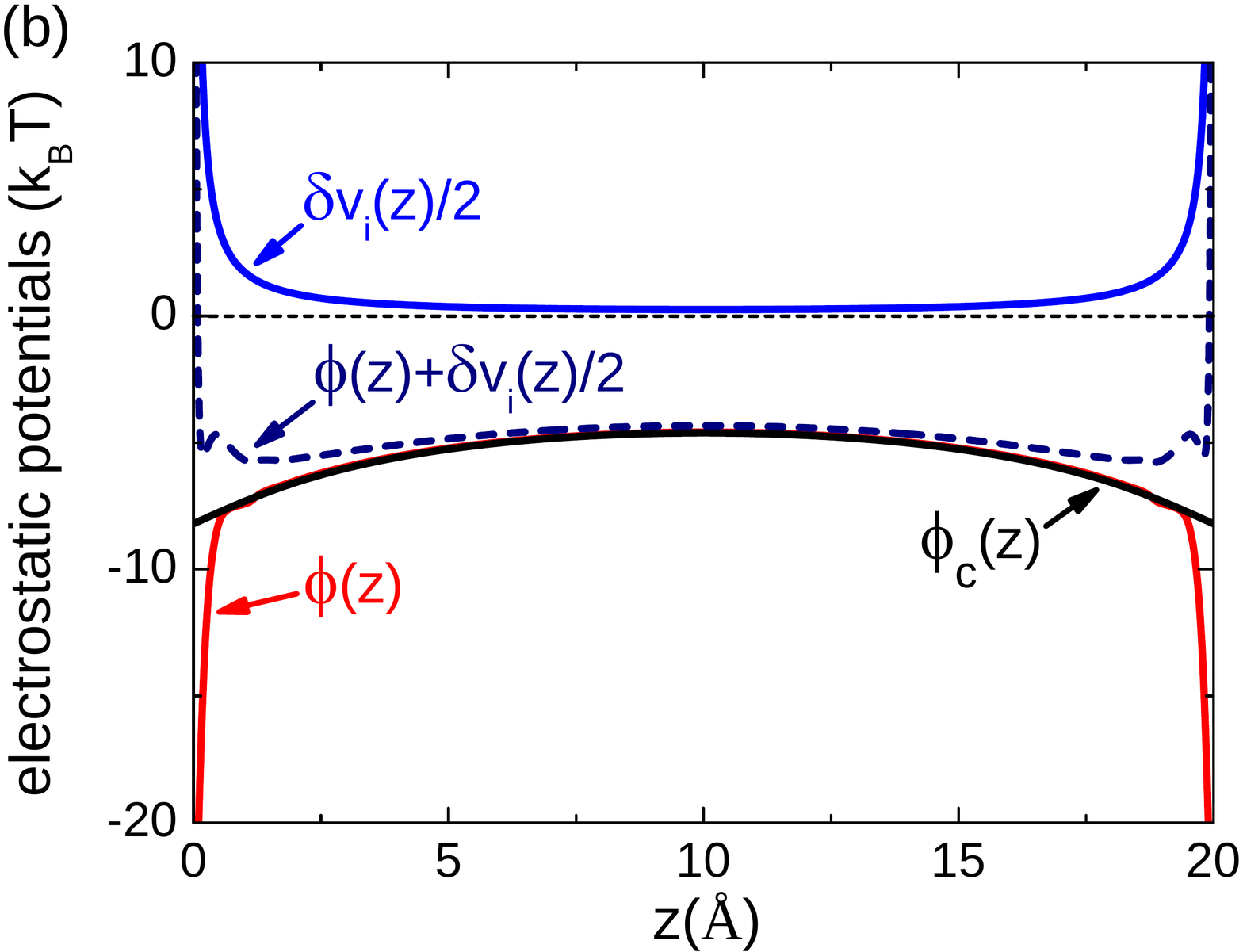}
\caption{(Color online) (a) Rescaled counterion densities in the slit of size $d=2$ nm and different charges increasing from bottom to top : $\sigma_s=0.7$ e $\mbox{nm}^{-2}$ (black curve), $1.0$ e $\mbox{nm}^{-2}$ (navy curve), and $1.3$ e $\mbox{nm}^{-2}$ (red curve). The inset zooms on the interfacial region. We also display the counterion density from the solvent-implicit theory (dashed navy curve). (b) Image charge potential $\delta v_i(z)$ (blue curve), electrostatic potentials $\phi_c(z)$ from the implicit solvent formalism (black curve) and $\phi(z)$ of the explicit solvent approach (red curve) at the pore charge $\sigma_s=1.0$ e $\mbox{nm}^{-2}$. Reservoir ion density and membrane permittivity are $\rho_{ib}=0.01$ M and $\e_m=1$ in both figures.}
\label{fig5}
\end{figure}

\subsection{Solvent effects on counterion partition}

In this part, we probe solvent structure effects on the partition of counterions in the slit pore. We illustrate in Fig.~\ref{fig5}(a) cation densities from the solvent-explicit formulation (solid curves) for increasing pore charge from bottom to top. We also reported the cation density profile from the continuum formulation of Eq.~(\ref{eq13}) (dashed curve at the surface charge $\sigma_s=1.0$ e $\mbox{nm}^{-2}$).  Fig.~\ref{fig5}(b) displays in turn the electrostatic potentials $\phi(z)$ and $\phi_c(z)$ as well as the ionic self-energy $\delta v_i(z)/2$ and the total ionic potential of mean force (PMF) $\phi(z)+\delta v_i(z)/2$.  We also note that in this figure, the strong amplitude of the repulsive ionic self-energy is known to result from image charge interactions between ions and the membrane of low dielectric permittivity. 

In Fig.~\ref{fig5}(a), the result from the continuum theory at the surface charge $\sigma_s=1.0$ e $\mbox{nm}^{-2}$ (dashed navy curve) shows that due to the competition between the repulsive image-charge potential and the attractive electrostatic potential, the density curve exhibits at both interfaces a counterion depletion over $1$ {\AA} followed by a maximum. The maxima are separated in turn by a well in the mid-pore area. At the same surface charge, the result of the solvent-explicit theory exhibits the same counterion peak, with the corresponding curve exhibiting fluctuations around the continuum result (see the inset). Most importantly, at the surface charge $\sigma_s=0.7$ e $\mbox{nm}^{-2}$, another counterion peak located at a closer distance from the interfaces appears. Increasing the surface charge from this value to $\sigma_s=1.3$ e $\mbox{nm}^{-2}$, the height of this peak rises and exceeds that of the second peak. The additional interfacial counterion adsorption layer absent in the continuum theory is a consequence of the non-uniform solvent configuration. Indeed, Fig.~\ref{fig5}(b) shows that due to the interfacial dielectric screening deficiency discussed in Section~\ref{partch},  the potential of the solvent-explicit formalism $\phi(z)$ is much larger than the continuum potential $\phi_c(z)$ in the interfacial regions. More precisely, at the surface charge  $\sigma_s=1.0$ e $\mbox{nm}^{-2}$,  the surface potential values are $\phi_c(0)\approx-8$ $k_BT$ and $\phi(0)\approx-27$ $k_BT$. In other words, close to the pore walls, the amplitude of the surface charge induced electrostatic force becomes comparable with image-charge forces. The location of the adsorption peaks absent in the continuum approach corresponds precisely to the region where the repulsive image-charge force is locally dominated by the strongly attractive electric field. We emphasize that the presence of this interfacial counterion adsorption layer may have important consequences in the functioning of nanofluidic devices~\cite{netznano1,netznano2}. We plan to explore this issue in an upcoming article.

\begin{figure}
\includegraphics[width=1.2\linewidth]{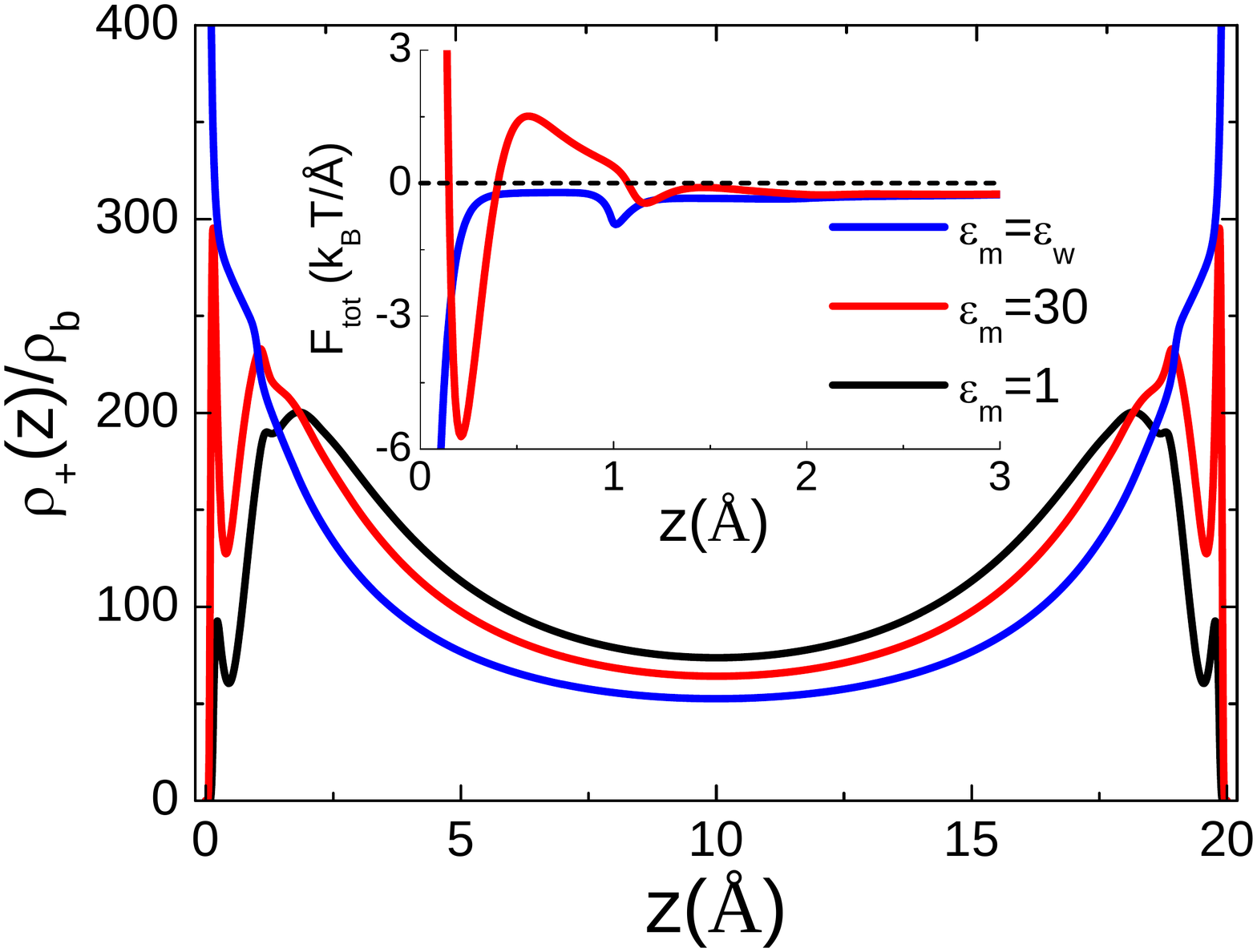}
\caption{(Color online) Counterion densities in the slit of size $d=2$ nm and surface charge $\sigma_s=0.7$ e $\mbox{nm}^{-2}$ for various membrane permittivities : $\e_m=1$ (black curve), $\e_m=30$ (red curve), and $\e_m=\e_w$ (blue curve). The inset shows the net electrostatic force $F_{tot}(z)=-\partial_z\left[\phi(z)+\delta v_i(z)/2\right]$ acting on each counterion.}
\label{fig6}
\end{figure}

We scrutinize now the effect of the membrane polarity on the ionic structure formation  illustrated in Fig.~\ref{fig5}(a). To this aim, we plotted in Fig.~\ref{fig6} counterion densities at three different membrane permittivities. The main plot shows that increasing the membrane permittivity from $\e_m=1$ to $30$,  the reduction of image charge forces amplifies the height of the first counterion peak. Because the electroneutrality condition fixes the total number of ions in the pore, this amplified counterion adsorption is compensated by a density decrease in the mid-pore. Moreover, with a further increase of the membrane polarity to the value $\e_m=\e_w$ where image-charge forces vanish, the two counterion peaks merge and leave a shouldering located at a distance $\sim1$ {\AA} from each interface. In this limit where the ionic structure disappears, the counterion density drops monotonically from the interface towards the mid-pore. We note that this behaviour is characteristic of the MF counterion densities previously investigated in Ref.~\cite{NLPB1} for single charged planes. 

Finally, in terms of the change in the membrane permittivity, we will show that the ionic structure formation results from the competition between surface charge attraction and image-charge repulsion. To this aim, we plotted in the inset of Fig.~\ref{fig6} the electrostatic force $F_{tot}(z)=-\partial_z\left[\phi(z)+\delta v_i(z)/2\right]$ experienced by each counterion located in the nanoslit. One sees that at the membrane permittivity $\e_m=30$ (red curve), the net force switches several times from repulsive ($F_{tot}(z)>0$) to attractive ($F_{tot}(z)<0$). At the points where the net force vanishes ($F_{tot}(z)=0$), surface charge attraction compensates exactly image charge repulsion. This gives rise to the counterion peaks displayed in the main plot.  In the limit $\e_m=\e_w$ where structure formation disappears (blue curve), the net force is seen to be purely attractive in the nanoslit ($F_{tot}(z)<0$). This leads in turn to a monotonical counterion increase towards the wall.

\section{Conclusions}

In this work, we investigated the coupled effects of membrane charge, polarization forces, and non-uniform dielectric permittivity on the partition of solvent molecules and ions under nanoconfinement. We showed that the solvent configuration is characterized by a strong correlation between dipolar orientation and solvent number density. Namely, local solvent excess is usually accompanied with the alignment of solvent molecules perpendicular with the pore walls, whereas solvent exclusion layers are associated with dipolar alignment parallel with the membrane. We also found that even in strongly charged membranes, subnanometer pores are characterized by a reduced solvent density and dielectric permittivity. This indicates that solvent-implicit ion rejection models assuming the same pore and reservoir permittivities are not reliable in predicting the ionic selectivity of subnanometer pores~\cite{yar1,yar2}. Moreover, the non-uniform polarization field of solvent molecules resulting in an interfacial dielectric screening deficiency leads to a strong surface electric field. This results in a counterion adsorption peak absent in continuum theories~\cite{pre1}. The observation of this additional counterion layer associated with solvent structure is the main result of our work. This peculiarity may have important effects on nanofluidic ion transport~\cite{netznano1,netznano2}. 

The present work aimed at probing the equilibrium properties of confined electrolytes associated with solvent charge structure. In an upcoming work, we plan to explore the impact of the points investigated herein on nanofluidic transport. It is also important to point out the limitations of the present approach. First, we emphasize that our non-MF NLPB formalism is hybrid since we approximated the electrostatic propagator accounting for charge correlation effects with the solution of the continuum DH equation. The reason for this approximation is the absence of any recipies for the solution of the solvent-explicit non-local kernel equation derived in Ref.~\cite{SBRB2}.  Moreover, although our formalism accounts for the extended solvent charge structure, the associated scalar field theory is a continuum formulation since it does not include excluded volume effects and dielectric cavities associated with solvent molecules. Despite these limitations, our approach goes beyond the classical continuum theories of electrolytes since it is the only formalism that can qualitatively reproduce interfacial ion and solvent structure formation observed in MD simulations. The comparison of the emerging ion transport properties with experimental data is of course needed in order to check the accuracy of the present theory and the consequences of the above-mentioned limitations.

\smallskip
\appendix

\section{Derivation of the ionic and dipolar self-energies} 
\label{ap1}

We report in this appendix the ionic and dipolar self-energies in Eqs.~(\ref{eq2}) and~(\ref{eq3}),
\bea
\label{eqI}
&&\delta v_i(z)=v(z,z)-v^b(0)\\
\label{eqII}
&&\delta v_d(z,a_z)=v_d(z,a_z)-2v^b(0)+2v^b(a),
\eea
with the electrostatic Green's function and dipolar potential
\bea\label{eqIII}
v(\br,\br')&=&\int_0^\infty\frac{\mathrm{d}kk}{2\pi}J_0\left[k|\br_\pa-\br'_\pa|\right]\tv(z,z'),\\
\label{eqIV}
v_d(z,a_z)&=&\int_0^\infty\frac{\mathrm{d}kk}{2\pi}\left\{\tv(z,z)+\tv(z+a_z,z+a_z)\right.\\
&&\hspace{1.7cm}-2\tv(z,z+a_z)\left.J_0\left[k|a_\pa|\right]\right\}\nonumber,
\eea
and the Bessel function of the first kind $J_0(x)$~\cite{math}. In Ref~\cite{SBRB2}, we had derived a variational kernel equation for the electrostatic Green's function $v(\br,\br')$. At present, an analytical or numerical approach to solve this highly non-local and non-linear kernel equation is not available. Thus, we will approximate  the electrostatic Green's function in Eqs.~(\ref{eqI})-(\ref{eqII}) by the solution of the local DH equation in slit pores~\cite{netzvdw},
\bea\label{eqV}
\tilde v(z,z',k)&=&\frac{2\pi\ell_w}{p}\left\{e^{-p|z-z'|}\right.\\
&&+\frac{\Delta}{1-\Delta^2e^{-2pd}}\left[e^{-p(z+z')}+e^{p(z+z'-2d)}\right.\nonumber\\
&&\hspace{2cm}\left.\left.+2\Delta e^{-2pd}\cosh\left(p|z-z'|\right)\right]\right\},\nonumber
\eea
where we have introduced the Bjerrum length in water solvent $\ell_w=e^2/(4\pi\e_wkT)$ and the functions
\bea\label{eqVI}
p&=&\sqrt{\kappa^2+k^2}\\
\label{eqVII}
\Delta&=&\frac{\e_wp-\e_mk}{\e_wp+\e_mk},
\eea
with the DH screening parameter $\kappa^2=4\pi\ell_w\sum_iq_i^2\rho_{ib}$. By substituting into Eqs.~(\ref{eqI})-(\ref{eqII}) the Fourier transformed propagator~(\ref{eqV}), the explicit form of the ionic and dipolar self-energies read
\bea\label{eqVIII}
\delta v_i(z)&=&\ell_w\int_0^\infty\frac{\mathrm{d}kk}{p}\frac{\Delta}{1-\Delta^2e^{-2pd}}\\
&&\hspace{1cm}\times\left\{e^{-2pz}+e^{-2p(d-z)}+2\Delta e^{-2pd}\right\}\nonumber\\
\delta v_d(z,a_z)&=&\ell_w\int_0^\infty\frac{\mathrm{d}kk}{p}\frac{\Delta}{1-\Delta^2e^{-2pd}}F(z,a_z),
\eea
where we introduced the auxiliary function
\bea
\label{eqIX}
F(z,a_z)&=&e^{-2pz}+e^{-2p(d-z)}+e^{-2p(d-z-a_z)}\\
&&+e^{-2p(z+a_z)}+4\Delta e^{-2pd}\nonumber\\
&&-2J_0\left[k|a_\pa|\right]\left\{e^{-p(2z+a_z)}+e^{-p(2d-2z-a_z)}\right.\nonumber\\
&&\hspace{2.1cm}\left.+2\Delta e^{-2pd}\cosh(pa_z)\right\}.\nonumber
\eea

\section{Relaxation algorithm for the solution of the non-MF NLPB equation~(\ref{eq1}) in slit pores}
\label{ap2}

We present herein a numerical relaxation scheme for the solution of Eq.~(\ref{eq1}) in slit pores. The scheme will be an extension of the algorithm presented in Ref.~\cite{SBRB2} for simple interfaces to the case of slit pores. In order to simplify the form of Eq.~(\ref{eq1}), we note that we considered in the present work the case of solvent molecules with monovalent elementary charges $Q=1$ and symmetric electrolytes composed of two monovalent ion species, i.e. $q_+=q_-=1$, with the bulk ion densities $\rho_{+b}=\rho_{-b}=\rho_{ib}$. With these simplifications, the NLPB Eq.~(\ref{eq1}) takes the form
\bea\label{eqX}
&&\partial_z^2\phi(z)-\e_w\kappa^2\;e^{-\delta v_i(z)/2}\sinh\left[\phi(z)\right]\\
&&-\kappa_s^2\int_{a_1(z)}^{a_2(z)}\frac{\mathrm{d}a_z}{2a}e^{-\delta v_d(z,a_z)/2}\sinh\left[\phi(z)-\phi(z+a_z)\right]\nonumber\\
&&=4\pi\ell_B\sigma_s\left[\delta(z)+\delta(d-z)\right],\nonumber
\eea
where we introduced the solvent screening parameter $\kappa_s^2=8\pi\ell_B\rho_{sb}$. By integrating Eq.~(\ref{eqX}) in the vicinity of the interfaces at $z=0$ and $z=d$, one obtains the two boundary conditions required for the solution of this equation,
\bea
\label{eqXI}
\phi'(0)&=&4\pi\ell_B\sigma_s\\
\label{eqXII}
\phi'(d)&=&-4\pi\ell_B\sigma_s.
\eea

The relaxation algorithm is based on the discretization of Eq.~(\ref{eqXI}) on the $z$ axis. This is done on a discrete lattice located between $z=0$ and $z=d$. The lattice is composed of $2N+1$ mesh points with separation distance $\epsilon$. In order to express Eq.~(\ref{eqX}) on the lattice, we introduce the discretized coordinate $z_n=\epsilon(n-1)$ and potential $\psi_n=\phi(z_n)$ with $2N+1\leq n\leq1$. By introducing as well the discrete form of the Laplacian operator $\epsilon^2\phi''(z)=\psi_{n+1}+\psi_{n-1}-2\psi_n$, Eq.~(\ref{eqX}) can be expressed on the lattice as
\bea
\label{eqXIII}
\psi_n&=&\frac{1}{2}\left\{\psi_{n+1}+\psi_{n-1}-r\;e^{-\delta v_i(z_n)/2}\sinh\left(\psi_n\right)\right.\\
&&\left.-s\sum_{j=j_1(n)}^{j_2(n)}e^{-\delta v_d(z_n,a_z\to z_{j-n+1})/2}\sinh\left(\psi_n-\psi_j\right)\right\},\nonumber
\eea
with the parameters $r=\e_w\epsilon^2\kappa^2$, $s=\e_w\epsilon^3\kappa^2$, and the functions 
\bea
j_1(n)&=&\mathrm{max}(1,n-n_a+1)\\
j_2(n)&=&\mathrm{min}(2N+1,n+n_a-1), 
\eea
where the index $n_a$ is defined as $z(n_a)=a$. Eq.~(\ref{eqXIII}) will be solved with the boundary conditions~(\ref{eqXI})-(\ref{eqXII}) that can be expressed on the lattice as $\psi_0=\psi_1-4\pi\ell_B\sigma_s\epsilon$ and $\psi_{2N+2}=\psi_{2N+1}-4\pi\ell_B\sigma_s\epsilon$. We finally note that on the same discrete lattice, the solvent number density~(\ref{eq7}) takes the form
\bea
\rho_d(z_n)&=&\frac{\rho_{sb}\epsilon}{a}\sum_{j=l_1(n)}^{l_2(n)}e^{-\frac{1}{2}\delta v_d(z_j,a_z\to 2z_n-2z_j)}\;e^{\psi_{2n-j}-\psi_j},\nonumber\\
\eea
with the boundaries 
\bea
l_1(n)&=&\mathrm{max}(1,n-n'_{a}+1,2n-2N-1)\\
l_2(n)&=&\mathrm{min}(2n-1,n+n'_{a}-1,2N+1), 
\eea
and the index $n'_{a}$ defined as $z(n'_{a})=a/2$.

The relaxation algorithm consists in solving Eq.~(\ref{eqXIII}) by iterating an initial reference potential $\phi_r(z)$. As already explained in Ref.~\cite{SBRB2}, the complication stems from the fact that for the convergence to be achieved, the input function has to verify the same boundary conditions as the NLPB Eq.~(\ref{eqX}), i.e. the guess potential should obey Eqs.~(\ref{eqXI})-(\ref{eqXII}). These conditions are not verified by the local MF PB equation since the PB potential satisfies a different boundary condition $\phi'(0)=4\pi\ell_w\sigma_s$ at $z=0$. In other words, the latter does not take into account the dielectric void in the neighbourhood of the charged interfaces. 

In order to account for the interfacial dielectric screening deficiency absent in the PB formalism, we will reiterate the trick explained in Ref.~\cite{SBRB2} and extend the guess potential derived in this earlier work for simple interfaces to slit nanopores. Namely, neglecting the dipolar self-energy $\delta v(z,a_z)$ in Eq.~(\ref{eqX}) and linearizing this equation in terms of the potential $\phi(z)$, one finds that the resulting equation is reduced for $z\ll a$ and $\e_w\kappa^2\ll\kappa_s^2$ to $\phi''(z)-c^2\kappa_s^2\phi(z)=-4\pi\ell_B\sigma(z)$, where we introduced the geometric factor $c^2=1/2$  associated with the rotational penalty for dipoles in the vicinity of the rigid interfaces. The solution of this equation yields the electrostatic field in the form $\phi'(z)=4\pi\ell_B\sigma_se^{-c\kappa_sz}$. We now note that this field is also the solution of the non-uniform Poisson equation $\partial_z\e(z)\partial_z\phi(z)=-4\pi\ell_B\sigma(z)$, with the dielectric permittivity function given by $\e(z)=e^{c\kappa_sz}$. In Ref.~\cite{SBRB2}, this exponential law was shown to reproduce accurately the behaviour of the dielectric permittivity up to the characteristic distance $d_1=\ln(\e_w)/(c\kappa_s)\approx 0.3$ {\AA} where the non-local permittivity reaches the bulk value (see Fig.6(a) of Ref.~\cite{SBRB2}). Inspired by this observation, we approximate for pores with thickness $d>2d_1$ Eq.~(\ref{eqX}) by the following equation,
\be\label{eqXIV}
\partial_z\e(z)\partial_z\phi(z)-\e_w\kappa_c^2(z)\left[\phi_r(z)-\phi_0\right]=-4\pi\ell_B\sigma(z),
\ee
where $\phi_0$ is a uniform Donnan potential that will be determined from a variational minimization procedure. Furthermore, the piecewise ionic screening parameter reads $\kappa_c(z)=\kappa\theta(z-d_1)\theta(d-d_1-z)$, and the inhomogeneous dielectric permittivity function is given by
\bea
\e(z)&=&e^{c\kappa_sz}\theta(d_1-z)\theta(z)+\e_w\theta(z-d_1)\theta(d-d_1-z)\nonumber\\
&&e^{c\kappa_s(d-z)}\theta(z-d+d_1)\theta(d-z).
\eea
The solution of Eq.~(\ref{eqXIV}) satisfying the continuity of the potential $\phi(z)$ and the displacement field  $\e(z)\phi'(z)$ at $z=d_1$ and $z=d-d_1$ reads
\bea\label{eqXV}
\phi_r(z)&=&\phi_0-\left[A+\frac{4\pi\ell_B\sigma_s}{c\kappa_s}e^{-c\kappa_sz}\right]\theta(d_1-z)\theta(z)\\
&&-\frac{4\pi\ell_w\sigma_s}{\kappa}\frac{\cosh\left[\kappa(d/2-z)\right]}{\sinh\left[\kappa(d/2-d_1)\right]}\theta(z-d_1)\theta(d-d_1-z)\nonumber\\
&&-\left[A+\frac{4\pi\ell_B\sigma_s}{c\kappa_s}e^{-c\kappa_s(d-z)}\right]\theta(z-d+d_1)\theta(d-z),\nonumber
\eea
where we introduced the constant
\be\label{eqXVI}
A=4\pi\ell_w\sigma_s\left\{\frac{1}{\kappa}\coth\left[\kappa\left(\frac{d}{2}-d_1\right)\right]-\frac{1}{c\kappa_s}\right\}.
\ee
The solution scheme consists in injecting into Eq.~(\ref{eqXIII}) the guess potential~(\ref{eqXV}) at the first iterative step, using the output function as the new input potential at the next iterative step, and continuing the cycle until numerical convergence is obtained. 

Finally, we explain the variational determination of the Donnan potential $\phi_0$ that takes into account non-linearities neglected by the linear trial form~(\ref{eqXV}). The approach consists in injecting the guess potential~(\ref{eqXV}) into the part of the variational Grand potential that generates Eq.~(\ref{eq1}). Up to constant factors that do not depend on the variational parameter $\phi_0$, the variational functional to be optimized with respect to $\phi_0$ reads~\cite{SBRB2}
\be\label{eqXVII}
h[\phi_0]=-2\rho_{ib}\int_0^d\mathrm{d}z\;e^{-\delta v_i(z)/2}\cosh[\phi_r(z)]-2\sigma_s\phi_0.
\ee
By taking the derivative of Eq.~(\ref{eqXVII}) with respect to the Donnan potential $\phi_0$, one obtains 
\be\label{eqXVIII}
-2\rho_{ib}\int_0^d\mathrm{d}z\;e^{-\delta v_i(z)/2}\sinh[\phi_r(z)]-2\sigma_s=0.
\ee 
The integral equation~(\ref{eqXVIII}) can be easily solved with a standard dichotomy algorithm in order to obtain the numerical value of the Donnan potential.


\begin{thebibliography}{99}
\bibitem {rem} The Fortran code that solves Eq.~(\ref{eq1}) can be obtained by contacting the author.
\end{thebibliography}
\end{document}